\documentclass{article}
\usepackage{graphicx}
\usepackage[minnames=3,maxnames=4,style=authoryear,maxbibnames=99,giveninits=true,backend=biber]{biblatex}
\DeclareNameAlias{sortname}{last-first}
\DeclareFieldFormat[article, inbook, incollection, inproceedings, misc, thesis, unpublished]{title}{#1.}
\renewbibmacro{in:}{}

\DeclareFieldFormat[article]
{volume}{ {#1} } 
\DeclareFieldFormat[article, inbook, incollection, inproceedings, misc, thesis, unpublished]
{number}{\mkbibparens{#1}} 
\DeclareFieldFormat[article]
{pages}{#1}
\DeclareFieldFormat[inproceedings, incollection, inbook]
{pages}{#1}
\renewbibmacro*{volume+number+eid}{
  \printfield{volume}
  \printfield{number}
  \printunit{\addcolon}
}

\DeclareFieldFormat{journaltitle}{\mkbibemph{#1}\isdot}

\renewbibmacro*{journal+issuetitle}{%
  \usebibmacro{journal}%
  \setunit*{\addcomma\space}%
  \iffieldundef{series}
    {}
    {\newunit
     \printfield{series}%
     \setunit{\addspace}}%
  \usebibmacro{volume+number+eid}%
  \setunit{\addspace}%
  \usebibmacro{issue+date}%
  \setunit{\addcolon\space}%
  \usebibmacro{issue}%
  \newunit}

\begin{document}
\DeclareFieldFormat{journaltitle}{#1} 

\noindent\textbf{Superconducting Materials for Microwave Kinetic Inductance Detectors}\\
Benjamin A. Mazin\\

\section{Operational Principle}

Microwave Kinetic Inductance Detectors, or MKIDs~(\cite{2003Natur.425..817D,Zmuidzinas:2012kh}), are superconducting resonators suitable for a wide variety of photon and particle detectors.  The surface impedance of a superconducting film is complex~(\cite{Mattis:1958vo}): it has an imaginary part, due to the reactance of the Cooper pairs, that exist together with the familiar geometrical inductance in any superconducting circuit. The real part is due to loss created by quasiparticle excitations, which behave similar to normal electrons. At low temperatures, $T << T_{\textrm{c}}$, the quasiparticle density is exponentially small resulting in a virtually lossless film, even at microwave frequencies.

\begin{figure}
\begin{center}
\includegraphics[width=1.0\columnwidth]{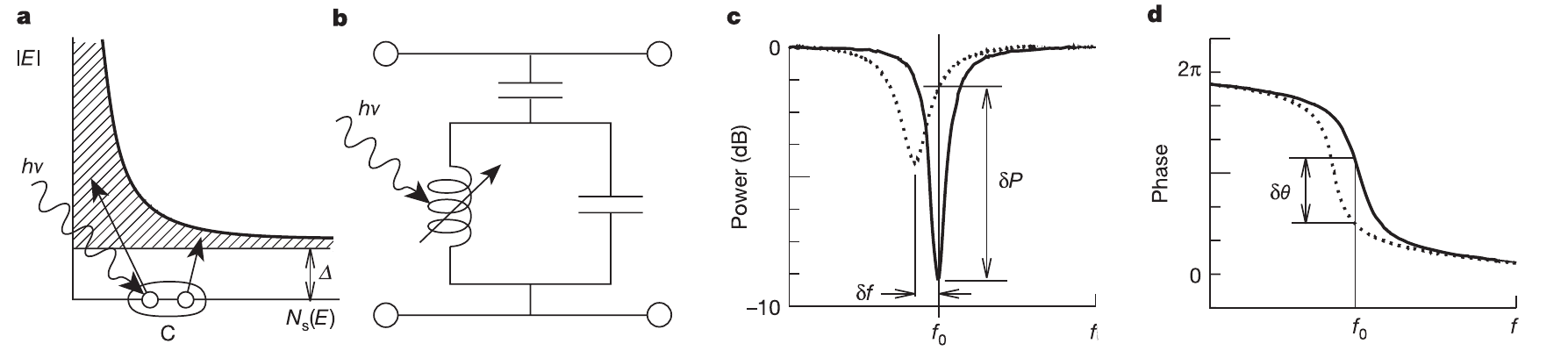}
\end{center}
\vspace{-0.2in}
\caption{The basic operational principles of an MKID. (a) Photons with energy $h\nu > 2 \Delta$ are absorbed in a superconducting film, producing a number of excitations, called quasiparticles.  (b) To sensitively measure these quasiparticles, the film is placed in a high frequency (MHz--GHz) planar resonant circuit.  The amplitude (c) and phase (d) of a microwave excitation signal sent through the resonator.  The change in the surface impedance of the film following a photon absorption event pushes the resonance to lower frequency and changes its amplitude.  If the detector (resonator) is excited with a constant on-resonance microwave signal, the energy of the absorbed photon can be determined by measuring the degree of phase and amplitude shift.  } 
\label{fig:detcartoon}
\end{figure}

This extra surface inductance changes when energy (which can come in the form of photons, phonons, or particles) breaks Cooper pairs into quasiparticles. The result is a change in complex surface impedance that depends on the number of Cooper pairs broken by incident photons, making it proportional to the amount of energy deposited in the superconductor. This surface impedance change can be measured by placing a superconducting inductor in a lithographed resonator, as shown in Figure~\ref{fig:detcartoon}.  More information on the kinetic inductance effect can be found in~\cite{Zmuidzinas:2012kh} and~\cite{Gao:2008td}. 

A MKID is essentially a microwave LC oscillator with a resonant frequency $f_\textrm{0}$ in the GHz range.  It has nearly perfect transmission away from the resonant frequency, but acts as a short on resonance, much like a high $Q$ notch filter.  More information on superconducting resonators can be found in chapter H.2.1.

\begin{figure}
\begin{center}
\includegraphics[width=0.6\columnwidth]{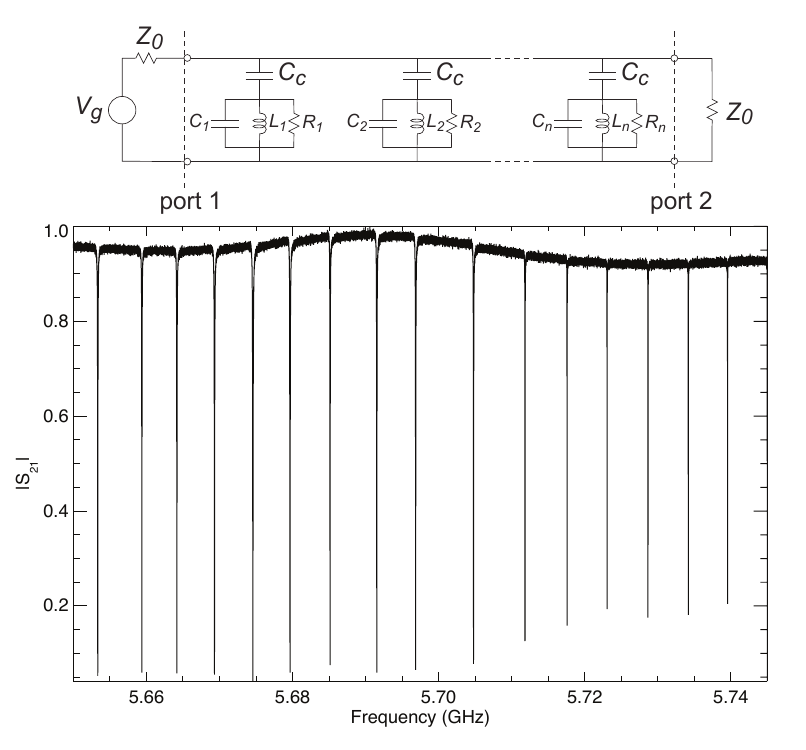}
\end{center}
\vspace{-0.2in}
\caption{An example of frequency domain multiplexing (FDM) of MKIDs.  Each MKID is a superconducting resonator tuned to a different resonant frequency by changing the resonator geometry.} 
\label{fig:fdm}
\end{figure}

To read out a MKID a microwave probe signal, typically a simple sine wave, is tuned to the $f_\textrm{0}$ of the resonator.  One photon (in the case of X-ray and optical MKIDs) or many photons (in the case of sub/mm MKIDs) imprint their signature as changes in phase and amplitude of this probe signal. Since the measured quality factor $Q_\textrm{m}$ of the resonators is high (usually 10,000--100,000) and the microwave transmission off resonance is almost perfect, multiplexing can be accomplished by tuning each pixel to a different resonant frequency during device fabrication.  A comb of probe signals (Figure~\ref{fig:fdm}) can be sent into the device, and room temperature electronics can recover the changes in amplitude and phase without significant cross talk~(\cite{2012RScI...83d4702M}).

The primary advantage over other superconducting detector technologies like transition edge sensors (TESs, ~\cite{Irwin:1996vk}) is that MKIDs are easy to multiplex, dramatically reducing the number of wires needed to read out an arrays, as shown in Figure~\ref{fig:fdm}.  They are also typically easier to fabricate than other superconducting sensors in that the usually do not use membranes or micromachining.

While the technology was originally envisioned for sub-mm detectors~(\cite{Doyle:2008gc,2016JLTP..184..816C,2016JLTP..184..173D,2017A&A...601A..89B}), MKIDs have found applications across the electromagnetic spectrum.  X-ray detectors using both non-equilibrium~(\cite{2006ApPhL..89v2507M}) and thermal detection~(\cite{Ulbricht:2015kz}) have been demonstrated, and significant progress has been made on optical and near-IR arrays for astronomy~(\cite{Mazin:2012kl,Szypryt:2017cb}).  Cosmic Microwave Background (CMB) detectors have also been a focus~(\cite{Hubmayr:2013gs,Matsumura:2016bh,McCarrick:2018cq}).  Works is also beginning on more serious attempts to use MKIDs for particle detection~(\cite{Moore:2012ja,Cruciani:2016et}).

\section{MKID Materials}

The superconducting materials that make up an MKID have a significant effect on its performance.  The $T_\textrm{c}$ and normal state resistivity $\rho_\textrm{N}$ of the film determine the penetration depth $\lambda$ and therefore how much kinetic inductance it has (see ~\cite{Zmuidzinas:2012kh}, Section 2.3).  The ratio of kinetic inductance to total inductance ($\alpha$), the volume of the inductor, and $Q_\textrm{m}$ determines the magnitude of the response to incoming energy (see ~\cite{Zmuidzinas:2012kh}, Section 5.3).  The quasiparticle lifetime $\tau_\textrm{qp}$ is the characteristic time during which the MKID's surface impedance is modified by the incoming energy. The response is fairly well understood, but it is only one part of determining the performance of a detector.  The noise is the other component, and it is far more complex.  The two main source of noise, amplifier noise and two-level system noise, are critically dependent on the maximum microwave readout power that can be used to probe the MKID (see~\cite{Zmuidzinas:2012kh}, Section 5.4).  The exact limits to maximum readout power are thought to be quasiparticle breaking from readout photons in aluminum, and the build up of non-linear kinetic inductance in higher resistivity superconductors.  More details can be found in~\cite{Zmuidzinas:2012kh}, Section 5.7.

Many materials have been explored for use in superconducting resonators and MKIDs, but that information is often not published or scattered around the literature.  This chapter contains information and references on the work that has been done with thin film lithographed circuits for MKIDs over the last two decades.  Note that measured material properties such as the internal loss quality factor $Q_\textrm{i}$ and quasiparticle lifetime $\tau_\textrm{qp}$ vary significantly depending on how the MKID superconducting thin film is made and the system they are measured in, so it is best to interpret all stated values as typical but not definitive.  Values are omitted in cases when there aren't enough measurements or there is too much disagreement in the literature to estimate a typical value.  In order to be as complete as possible some unpublished results from the author's lab are included and can be identified by the lack of a reference.  Unless noted all films are polycrystalline or amorphous.

\subsection{Conventional and High $T_\textrm{c}$ Materials: $T_\textrm{c}$=4--85~K }

High $T_\textrm{c}$ superconductors, for our purposes including conventional high $T_\textrm{c}$ materials and all low $T_\textrm{c}$ materials with transition temperatures from 4 to 85 K, are employed in MKID devices for three reasons.  First, they are often used to construct the microwave transmission lines that bring signals into the arrays because their lower surface impedances tend to make it easier to produce compact 50~$\Omega$ feedlines.  Second, they are sometimes used as the sensor material in the MKID for applications that have a high incident flux and/or require high temperature operation.  This higher temperature operation also reduces cost and complexity of the required cooling system.  An example of this kind of application is general purpose terahertz imaging~(\cite{2016RScI...87c3105R}).  Third, long wavelength MKID arrays usually use antennas or transmission lines to deliver the photons to the MKID sensor element.  To deliver photon without loss, the superconducting gap of the antenna/transmission line material must be larger than half of the photon energy $(\Delta > h\nu/2)$. For example, Nb has a cutoff frequency of 700 GHz while NbTiN can operate to ${\sim}$1.0 THz.

\begin{table}[htb!]
	\centering
	\begin{tabular}{|l|l|l|l|l|}
		\hline
		\textbf{Superconductor} & \textbf{$T_\textrm{c}$ (K)} & \textbf{$Q_\textrm{i}$} & \textbf{ $\tau_\textrm{qp}$ ($\mu$s)} & \textbf{Notes}\\
		\hline
		             YBCO &              85 &              4,000 &              & \\
		\hline
		             MgB$_2$ &         40      &         30,000      &          &  \\
		\hline
		             NbTiN &         15      &    $>10^6$           &      $<1$      &  \\
		\hline
		             MoRe &         8--13      &    $7 \times 10^5$           &         & Tunable  \\
		\hline
		             Nb &        9       &       $>5 \times 10^5$          &            &  \\
		\hline
		             Ta &        4.5       &      $>10^5$           &        30    &  \\
		\hline
	\end{tabular}
	\caption{Properties of High $T_\textrm{c}$ MKID Materials.  Tunable refers to compound materials whose $T_\textrm{c}$ can be adjusted by varying the stoichiometry.}
	\label{table:highTc}
\end{table}

\subsubsection{YBCO}

There has been limited work using resonators made out of the high $T_\textrm{c}$ superconductor YBCO for MKID work~(\cite{Lindeman:2014bs,2019ApPhL.114m2601C}), where YCBO forms both the resonator and feedlines.  While the measured internal quality factor $Q_\textrm{i}$ is low, on the order of 4,000, it is sufficient for detectors with high incident power, and operation at temperatures above 10 K has been demonstrated.

\subsubsection{MgB$_2$}

Some early work shows that MgB$_2$ is a possible material for MKIDs~(\cite{2018ApPhL.112b2601Y}).  The authors show operation at 7.5 K results in $Q_\textrm{i}\sim$30,000, despite the added complexity of the two-gap material.

\subsubsection{NbTiN}

NbTiN, with a $T_c$ of ${\sim}$15K, is commonly used as the ground plane material in MKID arrays~(\cite{Janssen:2013dg}). It is a well behaved and well understood material with low loss, and has found applications in similar devices like travelling wave parametric amplifiers~(\cite{HoEom:2012kq}).  Its low loss means that is occasionally used to define the capacitors in a MKID with different materials for the capacitor and the inductor, sometimes called a hybrid MKID.

\subsubsection{MoRe}

Preliminary work on the disordered superconducting alloy MoRe~(\cite{2014ApPhL.105v2601S}) has shown promising resonator internal quality factors.  It has potential applications as both a ground plane and resonator capacitor material, similar to NbTiN.

\subsubsection{Nb}

Nb is perhaps the most commonly used material for MKID ground planes as its elemental nature and wide use makes it almost a requirement in any superconducting fabrication line.  Nb is also used for antennas to bring signals to the sensor as well as in passive circuit elements like band-defining filters.  Nb resonators tend to have relatively high Q$_i >> 2 \times 10^5$~(\cite{Gao:2011ij,Thakur:2017dx}), although deposition conditions (low base pressure during sputtering, stress, and film purity) and substrate appear to play an important role.  Advanced Nb fabrication methods developed for superconducting electronics, such as planarization, enable very complex wiring~(\cite{2019ITAS...2902530D}).

\subsubsection{Ta}

Ta is often used as the absorber in X-ray MKIDs~(\cite{2006ApPhL..89v2507M}) because of its excellent X-ray stopping power due to its high atomic number (73) and density (16.65 g/cm$^3$).  Ta resonators have been fabricated and display similar microwave properties to Nb resonators, although Nb's higher $T_\textrm{c}$ means it is used more widely. The quasiparticle lifetime has been measured in several different ways to be approximately 30~$\mu$s~(\cite{2006ApPhL..89v2507M,2008PhRvL.100y7002B}).  Most Ta films are polycrystalline, but it has been successfully grown epitaxially by doing a heated deposition on r-plane sapphire.  While epi-Ta films have shown residual resistance ratios (RRR) greater than 30, the microwave loss properties appear to be similar to amorphous Ta~(\cite{2006ApPhL..89v2507M}).

\subsection{Low $T_\textrm{c}$ Materials: T$_c=0.8-4$~K }

Low $T_\textrm{c}$ superconductors, with transition temperatures from 0.8--4.5 K, are the most common materials used for the sensor element (the inductor of the resonator) in modern MKID arrays.  MKIDs are typically operated at $T_{\textrm{op}} \leq T_{\textrm{c}}$/8, so these materials are right in the sweet spot for operation with portable cryogenic technologies like He3 sorption coolers ($T_\textrm{op} \approx 300$ mK) and adiabatic demagnetization refrigerators (ADRs, $T_\textrm{op} \approx 100$ mK).

\begin{table}[htb!]
	\centering
	\begin{tabular}{|l|l|l|l|l|}
		\hline
		\textbf{Superconductor} & \textbf{$T_\textrm{c}$ (K)} & \textbf{$Q_\textrm{i}$} & \textbf{ $\tau_{qp}$ ($\mu$s)} & \textbf{Notes}\\
		\hline
		 Re & 1.4 & $>4 \times 10^6$ &  & \\
		\hline
		 Al & 1.2 & $>10^6$ & 2000 & \\
		\hline
		 Mo & 0.9 & $\sim$0 & & Normal metal oxides\\
		\hline
		 TiN & 0.8--4.5 & $>10^7$ & 50 & Tunable \\
		\hline
		 Ti/TiN & 0.8--4.5 & $>10^5$ & & Tunable \\
		\hline
		 WSi$_x$ & 1.4--4.5 & $>10^6$ & $<$1 & Tunable, high heat capacity \\
		\hline
		 PtSi & 0.93 & $>5 \times 10^5$ & 25 & Low $Q_\textrm{i}$ on Si\\
		\hline		
	\end{tabular}
	\caption{Properties of Low $T_\textrm{c}$ MKID Materials.}
	\label{table:lowTc}
\end{table}

\subsubsection{Re}

Rhenium~(\cite{Wang:2009ja,2016ITAS...2647221D}) has been explored as an alternative to aluminum, primarily because it has a similar $T_\textrm{c}$ but can be grown epitaxially on a sapphire substrate. Initial efforts showed inferior RF loss performance compared with Al and little effect from epitaxial growth.  It has not been pursued much farther because it is less common and more expensive than aluminum.

\subsubsection{Al}

Aluminum was the first MKID material~(\cite{2003Natur.425..817D}), and remains one of the most popular for both sensors and feedlines.  Aluminum films are easy to produce, have excellent uniformity, low loss, and long quasiparticle lifetimes.  The primary drawback of Al is that its low surface impedance often pushes the user to very thin films, which increases loss and $T_\textrm{c}$ (up to 2 K for the thinnest films) and makes them vulnerable to damage in processing.

Aluminum remains the technology of choice for CMB~(\cite{McCarrick:2018cq}) and sub-mm MKIDs~(\cite{Janssen:2013dg}), and is used widely in lithographed thin film resonators for quantum information experiments~(\cite{Megrant:2012cd,2014Natur.508..500B}).  It tends not to be used for shorter wavelength detectors where smaller pixels are required.

Significant effort has been spent doping aluminum with mangenese to lower its $T_\textrm{c}$.  This approach works and allows a relatively high $Q_\textrm{i}$ to be maintained~(\cite{2017ApPhL.110v2601J}) down to at least T$_c~\sim600$~mK, although it has been shown to lower the quasiparticle lifetime~(\cite{2008PhRvL.100y7002B}).  

Recent work has focused on using high resistivity granular aluminum (grAl) for resonators for both MKIDs and qubits~(\cite{2018PhRvL.121k7001G}).  It behaves mostly like aluminum but the sheet resistance can be tuned up to at least $4 \times 10^3$~$\mu \Omega$~cm, and likely higher as it approaches the superconductor/insulator transition.  It is relatively easy to make by depositing aluminum in the presence of oxygen, and is a good candidate when the simplicity of aluminum is helpful but a higher surface impedance per square is required.


\subsubsection{Mo}

Molybdenum resonators seem attractive due to the convenient $T_\textrm{c}$, but work at JPL failed to produce any working Mo resonators.  The most likely scenario is that molybdenum oxides are forming on the surface and remain conductive to low temperatures~(\cite{Huang:2017ge}).  These lossy conductive oxides mean that Mo is likely a material to avoid.

Some success has been had with Mo$_2$N~(\cite{2013ITAS...2300404P,Cataldo:2016dq}) with Q$_i >$~100,000.

\subsubsection{TiN$_x$}

Titanium Nitride is an excellent material for MKIDs as it has shown the lowest loss of any known material~(\cite{2010ApPhL..97j2509L}) and has a high normal state resistance which leads to high surface impedance and hence small, sensitive resonators.  It has a relatively long quasiparticle life for a disordered superconductor, and the $T_\textrm{c}$ can be tuned over a wide range (0.8--4.5 K) by adjusting the nitrogen content of the film.  The $T_\textrm{c}$ drops from stoichiometric TiN (4.5 K) to that of pure Ti (0.4 K) during reactive sputtering of TiN$_x$, where $x<1$. However, the extreme sensitivity of $T_\textrm{c}$ to nitrogen content has the effect of making the TiN film spatially non-uniform across the wafer due to process/gas flow imperfections, which shifts resonant frequencies around and lowers yield due to resonator collisions in frequency space.  Despite significant effort, low loss substoichiometric TiN$_x$ has only been grown on crystalline silicon substrates, while stoichiometric TiN may give good loss performance on other substrates.  

Some groups have had success growing more uniform stoichiometric titanium nitride using atomic layer deposition~(\cite{2013ITAS...23T0404C,szyprty15}).  Successful low loss ALD growth seems to involve making sure the background partial pressure of oxygen is as low as possible.  Making sub-stoichiometric TiN$_x$ with lower $T_\textrm{c}$ remains a challenge for ALD.

TiN$_x$ was used in the first generation of optical and near-IR (OIR) MKID arrays~(\cite{Mazin:2012kl}) and X-ray thermal kinetic inductance detectors (TKIDs)~(\cite{Ulbricht:2015kz}).  It is currently being, among other thins, for large area particle detectors~(\cite{2018JLTP..193..726C}).

\subsubsection{Ti/TiN Multilayers}

Due to the non-uniformity of reactively sputtered TiN a process was developed where alternating layers of Ti and stoichiometric 4.5 K TiN~(\cite{2013ApPhL.102w2603V}) were laid down in a superlattice. This leads to dramatically more uniform films, with preliminary measurements of NIST multilayers made at UCSB showing lower $Q_\textrm{i}$ when $T_\textrm{c}$ is pushed below 1 K. It is currently being used in feedhorn-coupled dual polarization submm/far-IR MKID arrays (\cite{2015ApPhL.106g3505H,2018JLTP..193..120A}). 

\subsubsection{WSi$_x$}

Tungsten Silicide is an interesting material with a $T_\textrm{c}$ that is tunable by adjusting the W to Si ratio~(\cite{2009JPhCS.150e2106K}). It has an extremely short quasiparticle lifetime, which made it interesting for the sensor in a Thermal Kinetic Inductance Detectors (TKIDs)~(\cite{Cecil:2012ds}).  Unfortunately recent work we did at UCSB has shown that WSi grown at UCSB has an anomalously high heat capacity at low temperatures, which degrades TKID performance significantly.

\subsubsection{PtSi}

\begin{figure}
\begin{center}
\includegraphics[width=0.4\columnwidth]{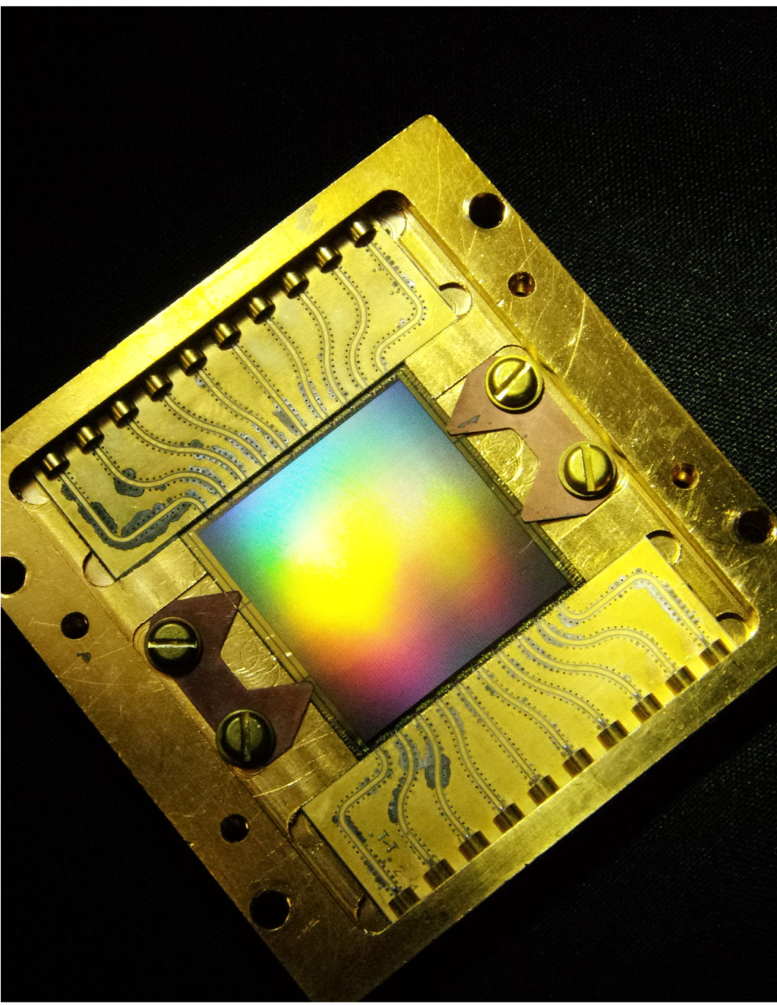}
\end{center}
\vspace{-.2in}
\caption{A 20,440 pixel MKID array optimized for near-IR astronomy from~\cite{Szypryt:2017cb}.} 
\label{fig:M1}
\end{figure}

The non-uniformity issues with TiN caused us to look for another superconducting material that would be stoichiometric at $T_\textrm{c} \sim 900$~mK, very uniform, and would work on an insulating substrate like sapphire.  We found this material in platinum silicide (PtSi), and it has led to the highest performance optical and near-IR MKID arrays produced, such as the array chip in a holder shown in Figure~\ref{fig:M1}~(\cite{Szypryt:2017cb}).  PtSi can be grown by depositing Pt on a Si wafer and doing a thermal anneal, but we observed that this leads to high loss as some Pt atoms diffuse deep into the Si creating a layer of normal metal under the superconducting PtSi~(\cite{Szypryt:2016ix}).  To avoid this problem and the issues present with OIR detectors on silicon~(\cite{2012SPIE.8453E..0BM,vanEyken:2015it}), we make low loss PtSi by depositing a layer of Si and a layer of Pt on a sapphire wafer and then doing a thermal anneal at a relatively low temperature (300--400 C). 

\subsection{Very Low $T_\textrm{c}$ Materials: T$_c<0.8$~K}

The next frontier in MKID sensor materials is going to very low $T_\textrm{c}$ materials in order to improve performance.  Performance is expected to improve since the signals get larger but the noise does not increase as much.  The quasiparticle lifetime $\tau_{\textrm{qp}} \propto 1 / T^3_{\textrm{c}}$~(\cite{KAPLAN:1976uq}) is also expected to increase.  However, care must be taken as microwave losses can increase when the readout frequency becomes a significant fraction of the gap frequency~\cite{2014PhRvL.112d7004D}.  This work also requires expensive dilution refrigerators, and significant care must be taken to thermalize the resulting devices.

\begin{table}[htb!]
	\centering
	\begin{tabular}{|l|l|l|l|l|}
		\hline
		\textbf{Superconductor} & \textbf{$T_\textrm{c}$ (K)} & \textbf{$Q_\textrm{i}$} & \textbf{ $\tau_{qp}$ ($\mu$s)} & \textbf{Notes}\\
		\hline
		  $\beta$-Ta & 0.6  &  &  & \\
		\hline		
		  Os & 0.66 (0.71-0.77)& $8 \times 10^4$ & 120 &  \\
		\hline
		  Ti & 0.40 & $<10^4$ &  & \\
		\hline
		  Hf & 0.13 (0.25-0.45) & $6 \times 10^5$ & 80 & \\
		\hline
		  Ir & 0.11 &  &  & No Oxides\\
		\hline
	\end{tabular}
	\caption{Properties of Very Low $T_\textrm{c}$ MKID Materials.  Measured $T_\textrm{c}$ of thin films shown in parenthesis.}
	\label{table:vlowTc}
\end{table}

\subsubsection{Ta ($\beta$ phase)}

Under certain conditions tantalum can be grown in the $\beta$ phase~(\cite{1972TSF....14..333S,Colin:2017be}), which lowers $T_\textrm{c}$ from 4.5 K to approximately 0.6 K.  We are not aware of any use of $\beta$-Ta for MKIDs, but it remains an intriguing possibility.

\subsubsection{Os}

Osmium is the densest material in the periodic table, and transitions around 0.66 K.  It has the fascinating property that it can be etched with oxygen, although the resulting Osmium tetroxide is a highly toxic gas.  We have successfully made sputtered Os resonators at UCSB.  However, due to concerns about toxicity and stability of devices in the atmosphere we have not pursued it. We measured Q$_i=$15,000 for ALD Osmium made by Jani Hamalainen in Helsinki, and 80,000 for sputtered Os made at UCSB.  We also measured a surprisingly long quasiparticle lifetime of 120~$\mu$s in the ALD Osmium devices.

\subsubsection{Ti}

Significant work was done at JPL to produce high quality titanium resonators using UHV sputtering, including using silicon and sapphire substrates, heating during deposition, and capping the film with a protective superconducting layer to try to prevent oxidation.  None of these efforts met with success, with $Q_{\textrm{i}}$ always less than 10,000.  Whether this is due to lossy metallic oxides or the Ti gettering impurities during sputtering is unknown.  

\subsubsection{Hf}

\begin{figure}
\begin{center}
\vspace{-0.4in}
\includegraphics[width=0.5\columnwidth]{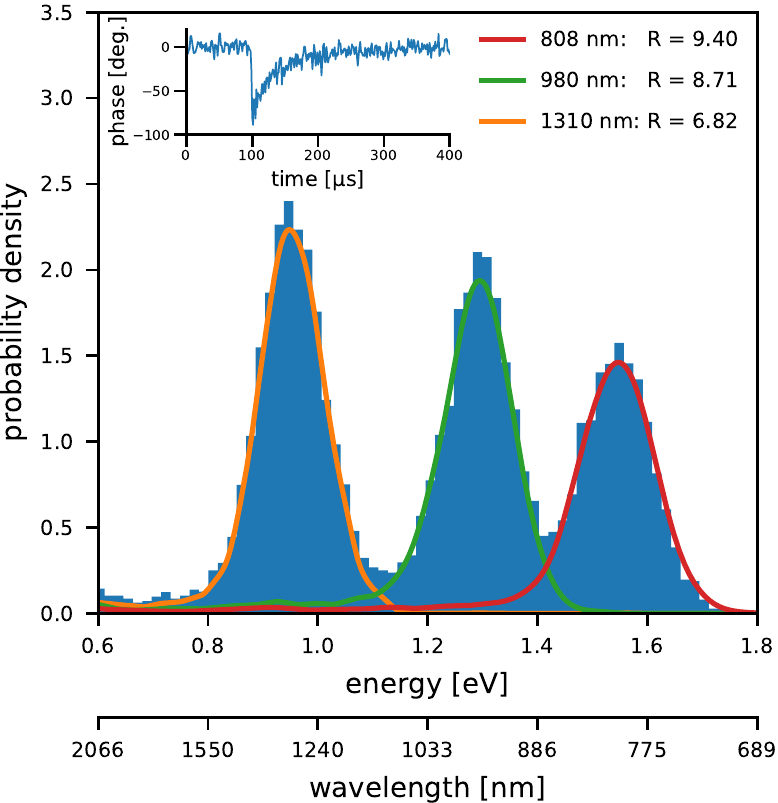}
\end{center}
\vspace{-.2in}
\caption{Spectral resolution of a single layer Hf resonator with $T_\textrm{c}$=395 mK, from \cite{2019ApPhL.115u3503Z}. Ordinate shows the probability of a photon arriving in this bin due to laser illunimation at 808, 980, and 1310 nm.  Despite being unoptimized, this Hf MKID is already producing superior energy resolution to PtSi. The poorly fit tail to longer wavelengths is due to the lack of a microlens, which allows photons to impact less sensitive areas of the resonator.} 
\label{fig:HfR}
\vspace{-.2in}
\end{figure}

We have been studying Hf resonators for MKID sensors at UCSB for the last several years~(\cite{2019ApPhL.115u3503Z,2020arXiv200400736C}).  The original Hf films we used were deposited at JPL, and more recently at both JPL and UCSB.  We have successfully deposited Hf on several different substrates, including high resistivity silicon and r, c, and a-plane sapphire. Initial devices had $T_\textrm{c}\sim450$~mK, but recent work at JPL has shown that $T_\textrm{c}\sim200$~mK can be achieved by changing the sputter pressure, as $T_\textrm{c}$ appears to depend on film stress.  In all cases the thin film $T_\textrm{c}$ has been significantly higher than the bulk $T_\textrm{c}=130$ mK.  $Q_\textrm{i}$ also appears to depend heavily on film stress with films with high compressive stress showing lower microwave loss.  UCSB films with  $T_\textrm{c}=$395 mK have a resistivity of 97 $\mu \Omega~\textrm{cm}$.  This yields a surface inductance of about 50 pH/sq for a 50 nm thick film, and 20 pH/sq for a 125 nm thick film.  Quasiparticle lifetimes in Hf appear to be relatively long, ${\sim}80~\mu$s.  Despite not being optimized, preliminary Hf MKIDs are already exceeding the spectral resolution of PtSi arrays (Figure~\ref{fig:HfR}).

\subsubsection{Ir}

Iridium is a non-reactive metal with a bulk $T_\textrm{c}$ around 112 mK.  Due to its non-reactive nature it cannot be etched with conventional reactive ion etches, and must be ion-milled to pattern.  JPL has seen resonances in Ir MKIDs, but has not characterized the material properties in a systematic way.

\section{MKID Instruments and Applications}

MKIDs are being used in a variety of astronomical and security applications, with more experiments coming online every year.  The following examples are illustrative of what is being done across a wide wavelength range.

\subsection{Optical/Near-IR}

MKID focal planes have so far been incorporated into two main classes of OIR instruments.  First, MKIDs have been used in seeing-limited integral field spectrographs (IFSs) for observations of time variable and low surface brightness objects with the ARCONS instrument~(\cite{2013PASP..125.1348M}) at the Palomar 200" telescope.  This testbed instrument led to a number of interesting scientific results~(\cite{2013ApJ...779L..12S,2014MNRAS.tmp..307S,Strader:2016he,2017ApJ...850...65C}), but a true facility class instrument is needed to make further progress. 

The second class of IFSs using MKIDs have been for high contrast direct imaging of exoplanets behind adaptive optics systems and a coronagraph.  For this application the near-IR sensitivity, lack of read noise, and fast readout enable active speckle suppression~(\cite{2014PASP..126..565M}) as well as extremely effective time-based speckle suppression using stochastic speckle discrimination~(\cite{Walter2019a}).  Existing MKID instruments for this application include DARKNESS for the Palomar 200"~(\cite{2018PASP..130f5001M}), MEC for Subaru SCExAO, and the PICTURE-C NASA balloon mission which is under construction.

\subsection{Sub-mm/CMB}

The MUSIC (\cite{2014SPIE.9153E..04S}), NIKA (\cite{2010A&A...521A..29M}), and NIKA2 (\cite{2018A&A...609A.115A}) instruments were some of the first instruments to use large arrays of MKIDs for mapping the sky in the sub-mm and mm wavelength. DESHIMA~(\cite{2019JATIS...5c5004E}) is novel sub-mm spectrometer using an on-chip filter bank spectrometer coupled to a MKID array.  It has recently been fielded at the ASTE telescope in Chile and returned its first astronomical results, included the detection of redshifted C[IV] emission from dusty star-forming galaxies~(\cite{2019NatAs.tmp..418E}).  Future iterations of DESHIMA and the similar instrument SUPERSPEC~(\cite{Wheeler:2018cg}) will be powerful tools for investigating the high redshift universe. 

\subsection{THz}

Airport security cameras can look though clothes and detect hidden objects.  A joint venture is building THz MKIDs designed for passive THz video imaging to faster and less intrusive airport screening~(\cite{2016RScI...87c3105R}).

\section{Discussion}

MKIDs are very flexible devices, but unfortunately new thin film MKID materials tend to be approached in a trial and error fashion due to the complexity of the possible outcomes and the large parameter space that must be explored, including deposition method (sputtering, e-beam or thermal evaporation, laser ablation, chemical vapor deposition, etc.), temperature, pressure, rate, and substrate.  Despite these challenges many suitable MKID materials have been discovered, allowing detector designers to chose materials most suitable for a given application.  In the coming years more materials will be explored, especially in the very low $T_\textrm{c}$ regime, paving the way for MKIDs of unprecedented sensitivity.

\printbibliography


\end{document}